\documentclass[journal=jpccck,manuscript=article,layout=twocolumn]{achemso}

\setkeys{acs}{
abbreviations=true,
articletitle=true,
biblabel=plain,
chaptertitle=true,
doi=true,
email=true,
etalmode=truncate,
keywords=true,
maxauthors=100,
super=true
}

\setcitestyle{super,open={},close={}}

\usepackage[utf8]{inputenc} 
\usepackage{textcomp} 
\usepackage[T1]{fontenc} 
\usepackage[portuguese,english]{babel} 
\usepackage[version=4]{mhchem} 
\usepackage{siunitx} 
\usepackage{graphicx} 
\graphicspath{{Figures/}} 
\usepackage{geometry} 
\usepackage[format=plain,justification=justified,singlelinecheck=false,font={normalsize,bf},labelfont=bf,labelsep=space]{caption} 
\usepackage{float} 
\usepackage{natbib} 
\usepackage{setspace} 
\usepackage{xkeyval} 
\usepackage{array} 
\usepackage{booktabs} 
\usepackage{listings} 
\usepackage{lmodern} 
\usepackage{mathpazo} 
\usepackage[breaklinks,colorlinks=true,allcolors=blue]{hyperref}
\hypersetup{colorlinks=true,citecolor=blue,urlcolor=blue}
\newcommand{\doi}[1]{\href{http://dx.doi.org/#1}{\nolinkurl{#1}}}
\usepackage{subfigure} 
\usepackage[compact]{titlesec} 
\usepackage{natmove} 
\usepackage{multirow} 
\usepackage{lipsum} 
\usepackage{calc} 
\usepackage{longtable} 
\usepackage{tabularx} 
\usepackage{placeins} 
\usepackage{tablefootnote}
\usepackage{adjustbox}
\usepackage{titletoc}
\usepackage{latexsym}
\usepackage{cancel}
\usepackage{amsmath}
\usepackage{amssymb}
\usepackage{amsfonts}
\usepackage{amstext}

\usepackage[none]{hyphenat}
\usepackage{microtype}
\makeatletter
\let\l@addto@macro\relax
\makeatother
\usepackage[fontsize=12pt]{scrextend}

\SectionsOn
\SectionNumbersOn
\AbstractOn

\title{Physicochemical Characterization of a New 2D Semiconductor Carbon Allotrope, C$_{\textbf{16}}$: An Investigation via Density Functional Theory and Machine Learning-based Molecular Dynamics}

\author{Kleuton A. L. Lima}
\affiliation[UNICAMP]{Department of Applied Physics and Center for Computational Engineering and Sciences, State University of Campinas, Campinas, São Paulo, 13083-859, Brazil.}

\author{Rodrigo A. F. Alves}
\affiliation[UnB]{Institute of Physics, University of Bras{\'{i}}lia, Bras{\'{i}}lia, 70910-900, Brazil}
\alsoaffiliation[LCCMat]{Computational Materials Laboratory, LCCMat, Institute of Physics, University of Bras{\'{i}}lia, 70910-900,
Bras{\'{i}}lia, Brazil}

\author{Elie A. Moujaes}
\affiliation[UNIR]{Physics Department, Federal University of Rondônia, 76801-974, Porto Velho, Brazil}

\author{Alexandre C. Dias}
\affiliation[UnB]{Institute of Physics, University of Bras{\'{i}}lia, Bras{\'{i}}lia, 70910-900, Brazil}
\alsoaffiliation[CIF]{International Center of Physics, University of Bras{\'{i}}lia, Bras{\'{i}}lia, 70910-900, Brazil}

\author{Douglas S. Galvão}
\affiliation[UNICAMP]{Department of Applied Physics and Center for Computational Engineering and Sciences, State University of Campinas, Campinas, São Paulo, 13083-859, Brazil.}

\author{Marcelo L. Pereira, Jr}
\affiliation[ENE]{University of Bras{\'{i}}lia, College of Technology, Department of Electrical Engineering, 70919-970, Bras{\'{i}}lia, Federal District, Brazil.}

\author{Luiz A. Ribeiro, Jr}
\email{ribeirojr@unb.br}
\affiliation[UnB]{Institute of Physics, University of Bras{\'{i}}lia, Bras{\'{i}}lia, 70910-900, Brazil}
\alsoaffiliation[LCCMat]{Computational Materials Laboratory, LCCMat, Institute of Physics, University of Bras{\'{i}}lia, 70910-900,
Bras{\'{i}}lia, Brazil}


\begin{document}

\maketitle


\begin{abstract}
This study comprehensively characterizes, with suggested applications, a novel two-dimensional carbon allotrope, C$_{16}$, using Density Functional Theory and machine learning-based molecular dynamics. This nanomaterial is derived from naphthalene and bicyclopropylidene molecules, forming a planar configuration with sp$^2$ hybridization and featuring 3-, 4-, 6-, 8-, and 10-membered rings. Cohesive energy of \SI{-7.1}{\electronvolt/atom}, absence of imaginary frequencies in the phonon spectrum, and the retention of the system's topology after ab initio molecular dynamics simulations confirm the structural stability of C$_{16}$. The nanomaterial exhibits a semiconducting behavior with a direct band gap of \SI{0.59}{\electronvolt} and anisotropic optical absorption in the $y$ direction. Assuming a complete absorption of incident light, it registers a power conversion efficiency of \SI{13}{\percent}, demonstrating relatively good potential for applications in solar energy conversion. The thermoelectric figure of merit ($zT$) reaches 0.8 at elevated temperatures, indicating a reasonable ability to convert a temperature gradient into electrical power. Additionally, C$_{16}$ demonstrates high mechanical strength, with Young's modulus values of \SI{500}{\giga\pascal} and \SI{630}{\giga\pascal} in the $x$ and $y$ directions, respectively. Insights into the electronic, optical, thermoelectric, and mechanical properties of C$_{16}$ reveal its promising capability for energy conversion applications.
\end{abstract}

\section{Introduction} 

The fabrication of new nanomaterials has always been associated with technological progress. Advancing nanotechnology requires a deep understanding of nanomaterials.\cite{suhag2023introduction} Nanomaterials can exist at different dimensionalities, from zero (quantum dots) to three dimensions (such as diamonds). In particular, two-dimensional (2D) materials have recently attracted significant attention due to their properties and potential applications in various technological fields,\cite{yadav2023synthesis} including electronics,\cite{alshammari2023organic} optoelectronics,\cite{lipovka2024laser} and energy storage.\cite{yu2024nature} 

Since the experimental realization of graphene,\cite{novoselov2004electric} more than two decades ago, the investigation of new carbon allotropes and other 2D materials has been at the forefront of materials science. Graphene is a semimetal material \cite{rao2009graphene} with high tensile strength \cite{lee2008measurement} and elevated thermal stability \cite{wu2009synthesis}. These attractive properties have motivated the search for new materials with similar or improved properties.\cite{perreault2015environmental}

Among the growing number of 2D materials families, carbon allotropes are particularly promising due to carbon's ability to form multiple bonding configurations (sp, sp$^2$, and sp$^3$ hybridizations and combinations of them),\cite{peschel2011carbon} resulting in a large diversity of physical characteristics dependent on the structural topology of atomic arrangements.\cite{tiwari2016magical} Although graphene exhibits high electrical conductivity, it has a zero electronic bandgap, limiting some kinds of applications.\cite{chang2013graphene} This limitation has stimulated research into new 2D carbon allotropes that combine graphene's structural benefits with semiconducting properties.\cite{nguyen2016two}

This growing interest in carbon-based 2D systems has led to the discovery and/or proposal of several new allotropes with distinct characteristics. Although most of these structures are from theoretical predictions \cite{tiwari2016magical,falcao2007carbon,zhang2019art,longuinhos2014theoretical,brandao2023mechanical,tromer2024mechanical}, significant advances have recently been made in the synthesis routes of these nanomaterials. 

In addition to graphene,\cite{novoselov2004electric} fullerenes,\cite{kroto1985c60}, and carbon nanotubes,\cite{iijima1993single}, which were synthesized years ago, recent experimental realizations include Biphenylene Network (BPN),\cite{fan2021biphenylene} Holey Graphyne (HGY),\cite{liu2022constructing} $\gamma$-graphyne ($\gamma$-GY),\cite{li2018synthesis,desyatkin2022scalable,he2023one} and monolayer fullerene Network (2D-C$_{60}$) \cite{hou2022synthesis}. Importantly, these systems were theoretically predicted before their synthesis \cite{baughman1987structure,balaban1968chemical,berber2004rigid,hudspeth2010electronic} 

Among these 2D carbon-based systems, graphene and BPN exhibit metallic behavior, while the HGY, 2D-C$_{60}$, and $\gamma$-GY monolayers are semiconductors.

In our quest to discover new 2D carbon allotropes, our groups have recently characterized 30 two-dimensional structures with semiconducting properties, specifically focusing on their solar energy conversion capabilities. Our results demonstrate that power conversion efficiency (PCE) values range from \SIrange{7}{30}{\percent}, with the possibility of complete absorption of an incident light.\cite{Dias_2024} Beyond their main use in electronic devices, these 2D semiconducting carbon allotropes also appear to be promising candidates for renewable energy applications.\cite{tang2014two}

In this context, in this work, we propose a new carbon allotrope named C$_{16}$. Density Functional Theory (DFT) simulations were used to characterize its dynamical and thermal stability, electronic structure, and optical, excitonic, and thermoelectric properties. 

We have also performed classical molecular dynamics (MD) simulations to study its mechanical properties. For the description of the inter-atomic interactions, we have employed a reactive force field trained using Machine Learning protocols based on the data obtained from ab initio Molecular Dynamics (AIMD) simulations for representative structures related to C$_{16}$. 

Our results show that C$_{16}$ is dynamically and structurally stable, with no imaginary frequencies in its phonon spectrum. It is also thermally stable, as no phase changes were observed when subjected to a thermal bath at \SI{1000}{\kelvin}. C$_{16}$ exhibits a semiconducting behavior, with a direct electronic bandgap of \SI{0.59}{\electronvolt}. Regarding its optical properties, C$_{16}$ exhibits anisotropic characteristics, with the main absorption activity occurring along one specific direction of the system. It has an exciton binding energy of approximately \SI{96}{\milli\electronvolt}, a power conversion efficiency (PCE) of \SI{13}{\percent}, and a figure of merit of 0.8. Regarding its mechanical properties, our results suggest an anisotropic behavior, with Young's modulus ranging from \SIrange{500}{630}{\giga\pascal}, depending on the direction of deformation.

\section{Methodology}

To better present the selected computational methods and simulation parameters, we have divided this section into three subsections: the modeling of C$_{16}$, the first-principles calculations, and the aspects of classical MD simulations.

\subsection{System Modeling}

\begin{figure*}[t!]
    \centering
    \includegraphics[width=16cm]{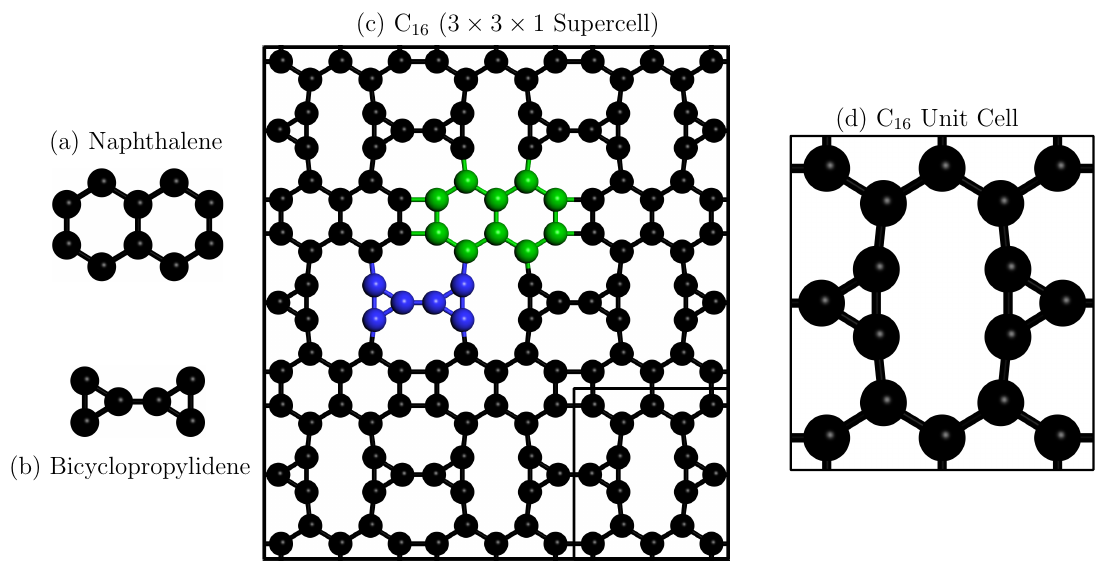}
    \caption{Schematic representation of the proposed C$_{16}$ structure, showing the molecules of (a) Naphthalene and (b) Bicyclopropylidene, which can be considered as the C$_{16}$ structural building blocks. (c) a $3\times 3\times 1$ C$_{16}$  supercell, with the molecular motifs highlighted in green and blue. (d) C$_{16}$ structural unit cell.}
    \label{fig:system}
\end{figure*}

The proposition of new nanomaterials based on carbon or other systems can have different approaches. A common approach, also used frequently in experimental processes, is to start with known hydrocarbon molecular motifs and dihydrogenated them to induce reactions.

In the literature, several structures have been obtained through the aforementioned process. Examples include the BPN and 2D-C$_{60}$ networks, which are obtained through the covalent bonding of BPN and C$_{60}$ molecules, respectively.\cite{fan2021biphenylene,hou2022synthesis} For $\gamma$-GY, several small molecules can be used, particularly dehydrobenzoannulenes (DBAs).\cite{li2023artificial} This approach has been used for several proposed carbon monolayers, such as PAI-G,\cite{chen2020pai} PSI-G,\cite{li2017psi} PCF,\cite{shen2019pcf} Naphthylene,\cite{beserra2020naphthylene} DHQ,\cite{wang2019dhq} TH-GY,\cite{lima2024th} Nanoporous Graphene,\cite{moreno2018bottom} among others.

In this perspective, here we propose a new allotrope, which consists of one possible 'fusion' of the naphthalene \cite{parker2012low} and bicyclopropylidene \cite{de2000bicyclopropylidene} into a 2D structure that is also planar (like the molecules), see Fig.~\ref{fig:system}. The molecular motifs are shown in Fig.~\ref{fig:system}(a-b). Naphthalene is an oligoacene with ten carbons, with the formula \ce{C10H8}, while bicyclopropylidene is a tetrasubstituted alkene with the formula \ce{C6H8}. 

The molecules were arranged in alternating linear blocks, forming covalent bonds. The system was designed to exhibit fully sp$^2$ hybridization. Fig.~\ref{fig:system}(c) shows a supercell of the system obtained from this process, with the molecular motifs highlighted in blue and green. The obtained structure (C$_{16}$) consists of carbon fused rings with 3, 4, 6, 8, and 10 atoms. The unit cell  (r\ce{C16}) contains 16 carbon atoms, Fig.~\ref{fig:system}(d).

\subsection{First Principles Calculations}

To characterize the physicochemical properties of C$_{16}$, we carried out first-principles simulations within the DFT formalism using the Vienna Ab initio Simulation Package (VASP).\cite{Kresse_13115_1993, Kresse_11169_1996} The Kohn-Sham (KS) equations were solved via the projector augmented wave (PAW) method,\cite{Kresse_1758_1999} applying a plane-wave cutoff energy of \SI{868.00}{\electronvolt}.  We used the Generalized Gradient Approximation (GGA) with the Perdew-Burke-Ernzerhof (PBE) exchange-correlation functional,\cite{perdew1996generalized} enforcing a convergence criterion of \SI{0.01}{\electronvolt/\angstrom} for minimizing atomic forces and the stress tensor.

After the geometry optimization (unit cell parameters and atomic positions), we obtained the C$_{16}$ electronic properties with a cutoff energy of \SI{488.23}{\electronvolt} adopting a total energy convergence criterion of \SI{E-6}{\electronvolt} to ensure self-consistency in the electronic density. Due to systematic errors associated with the PBE functional in estimating the electronic bandgap value \cite{Cohen_115123_2008, Crowley_1198_2016}, we applied the HSE06 hybrid exchange-correlation functional \cite{heyd_1187_2004, hummer_115205_2009} for a more realistic estimation of the electronic and optical gaps. To avoid spurious interactions between the monolayer and its periodic images along the $z$-direction, a vacuum buffer space of \SI{30}{\angstrom} was used.

In all calculations, the $k$-meshes were generated automatically using the Monkhorst-Pack method,\cite{monkhorst1976special} ensuring a density of \SI{40}{\per\angstrom} along the in-plane lattice vector directions,\cite{martin2020electronic} which results in a $6\times6\times1$ $k$-mesh for the unit cell. Dynamic stability was assessed by obtaining the phonon dispersion spectrum of a $3\times3\times1$ supercell with 64 atoms, with the same convergence criteria mentioned above. To evaluate the thermal stability, we performed AIMD simulations under the canonical NVT ensemble, using a Nosé-Hoover thermostat \cite{nose1984unified} to control the temperature value set at \SI{1000}{\kelvin}, with a time step of \SI{1}{fs} for a total simulation time of \SI{5}{ps}.

To investigate the excitonic and optical properties, we used a maximally localized Wannier function tight-binding (MLWF-TB) \cite{Arash_685_2008,Dias_108636_2023} Hamiltonian to solve the Bethe-Salpeter equation (BSE),\cite{Salpeter_1232_1951} due to the high computational cost of GW+BSE \textit{ab-initio} approaches.\cite{Deslippe_1269_2012,Sangalli_325902_2019} To determine MLWF-TB, through the Wannier90 code\cite{Arash_685_2008,pizzi2020wannier90,mostofi2014updated} we choose the \ce{C} s- and p-orbital projections, based on the electronic density of states results, complemented with a random basis to achieve the same number of calculated bands in the DFT simulations.

The optical properties were analyzed at both the BSE level, which includes excitonic effects, and the independent particle approximation (IPA), which does not take into account the quasi-particle effects. These analyses were carried out using the WanTiBEXOS code.\cite{Dias_108636_2023} The BSE was solved using a 2D-truncated Coulomb Potential (V2DT)\cite{Rozzi_205119_2006} and a  \textbf{k}-mesh of $19\times17\times1$; the highest two valence bands and the lowest three conduction bands were considered in the calculations to obtain an accurate description of the optical properties in the solar emission range (i.e. \SIrange{0.5}{4.0}{\electronvolt}). To ensure accurate results, we applied a smearing of \SI{0.01}{\electronvolt} to the dielectric functions at both the BSE and IPA levels.

The thermoelectric properties, namely the Seebeck coefficient, electronic conductivity, and thermal electronic conductivity, were computed using the Boltzmann code.\cite{pizzi2014boltzwann} This tool is interfaced with the VASP code within the Wannier90 package and depends on correctly producing the DFT-based band structure using MLWFs.\cite{kohn1973wannier,kohn1959analytic} Convergence is achieved by choosing a coarse  $24 \times 24 \times 1$ electronic grid for self-consistent calculations and a very dense $220 \times 220 \times 1$ electronic grid for the subsequent thermoelectric calculations.

\subsection{Classical Molecular Dynamics Simulations}

For the analysis of the C$_{16}$ mechanical properties, we adopted an advanced approach based on machine learning interatomic potentials (MLIP).\cite{novikov2020mlip} These potentials have been proven highly effective in overcoming the limitations of more standard methods, such as empirical interatomic potentials (EIP).\cite{mortazavi2020exploring} MLIPs enable the integration of \textit{ab initio} methods with classical MD simulations, allowing for a detailed investigation of mechanical responses and fracture mechanisms in diverse materials, using interatomic potentials obtained explicitly for the investigated-like systems.

Here, we developed a specific force field for C$_{16}$ based on the moment tensor potential (MTP),\cite{shapeev2016moment} trained with data generated from AIMD simulations in VASP. These molecular dynamics simulations were performed on supercells containing 64 atoms, covering uniaxial and biaxial strain changes to reduce the high correlation among the training data. Additionally, dynamics with progressive heating (heat ramps) were considered, resulting in \num{6500} training data points. This approach has been extensively validated in the literature, demonstrating success in predicting the mechanical properties of several nanomaterials.\cite{mortazavi2021first, mortazavi2023atomistic, lima2024th}

For MD simulations using the trained MLIP, we employed the LAMMPS software (Large-scale Atomic/Molecular Massively Parallel Simulator),\cite{plimpton1995fast}. We have used the Velocity-Verlet algorithm to integrate Newton's equations of motion,\cite{verlet1967computer} with a time step of \SI{0.05}{fs}. After optimizing the systems using the obtained MLIP, we carried out two additional steps to the structures for the tensile loading simulations. The release of any residual stresses was controlled through simulations in an isobaric-isothermal (NPT) ensemble for \SI{100}{ps}, using a Nosé-Hoover thermostat \cite{hoover1985canonical} to ensure thermal equilibrium and control. The temperature was set to \SI{300}{\kelvin} and the external pressure to zero. For further tests and/or comparisons, another simulation run for an additional \SI{100}{ps} was performed using a canonical ensemble (NVT) at ambient temperature. 

The elastic properties and fracture patterns were evaluated at \SI{300}{\kelvin}, applying a uniaxial strain rate of \SI{E-6}{\per\femto\second} along the $x$ and $y$ directions separately. The system was considered periodic in all directions, with pressure adjustments made perpendicular to the direction of the strain. The stress was calculated using the stress tensor for each atom, considering the virial contributions and atomic mass. Since the system is two-dimensional, the volume was defined based on the dimensions $L_x$ and $L_y$, adopting a thickness of \SI{3.35}{\angstrom}, consistent with the literature.\cite{Nair2008}

\section{Results and Discussion}

This section examines the physicochemical properties of C$_{16}$ based on the simulations using the abovementioned methodologies. We have explored its structural characteristics, including energy stability, dynamics, and thermal properties. Additionally, we investigated its electronic, excitonic, optical, thermoelectric, and mechanical properties. Potential applications of C$_{16}$, such as in solar cells, are also discussed.

\subsection{Structural Stability}

After the DFT geometry optimization, we observed that C$_{16}$ remains in a fully planar conformation (purely sp$^2$ hybridization). The symmetric triangular rings have one bond length of \SI{1.44}{\angstrom}, shared with the 10-membered ring. The other two equal sides of the triangular rings have bond lengths of \SI{1.38}{\angstrom}, shared with the 8-membered rings. The corresponding bond angle values are \SI{58.43}{\degree} and \SI{63.14}{\degree}, respectively. 

The 4-membered ring is a rectangle with one side with a bond length of \SI{1.41}{\angstrom} (shared with the hexagonal rings) and the other with \SI{1.54}{\angstrom} (shared with the octagonal rings). The bond angle values of the benzene rings vary from \SIrange{117.5}{120.8}{\degree}117.5. For the octagonal rings, the bond angle values range from \SIrange{115.4}{148.4}{\degree}, while for the decagon ring, the values vary from \SIrange{118.4}{173.7}{\degree}. The C$_{16}$ lattice vectors are \SI{6.48}{\angstrom} and \SI{7.15}{\angstrom} for the horizontal and vertical directions, respectively. C$_{16}$ belongs to the \textit{Pmmm} (D$_{\rm{2h}}^{\rm{1}}$) symmetry group.

\begin{figure*}[!htb]
	\centering
	\includegraphics[width=0.7\linewidth]{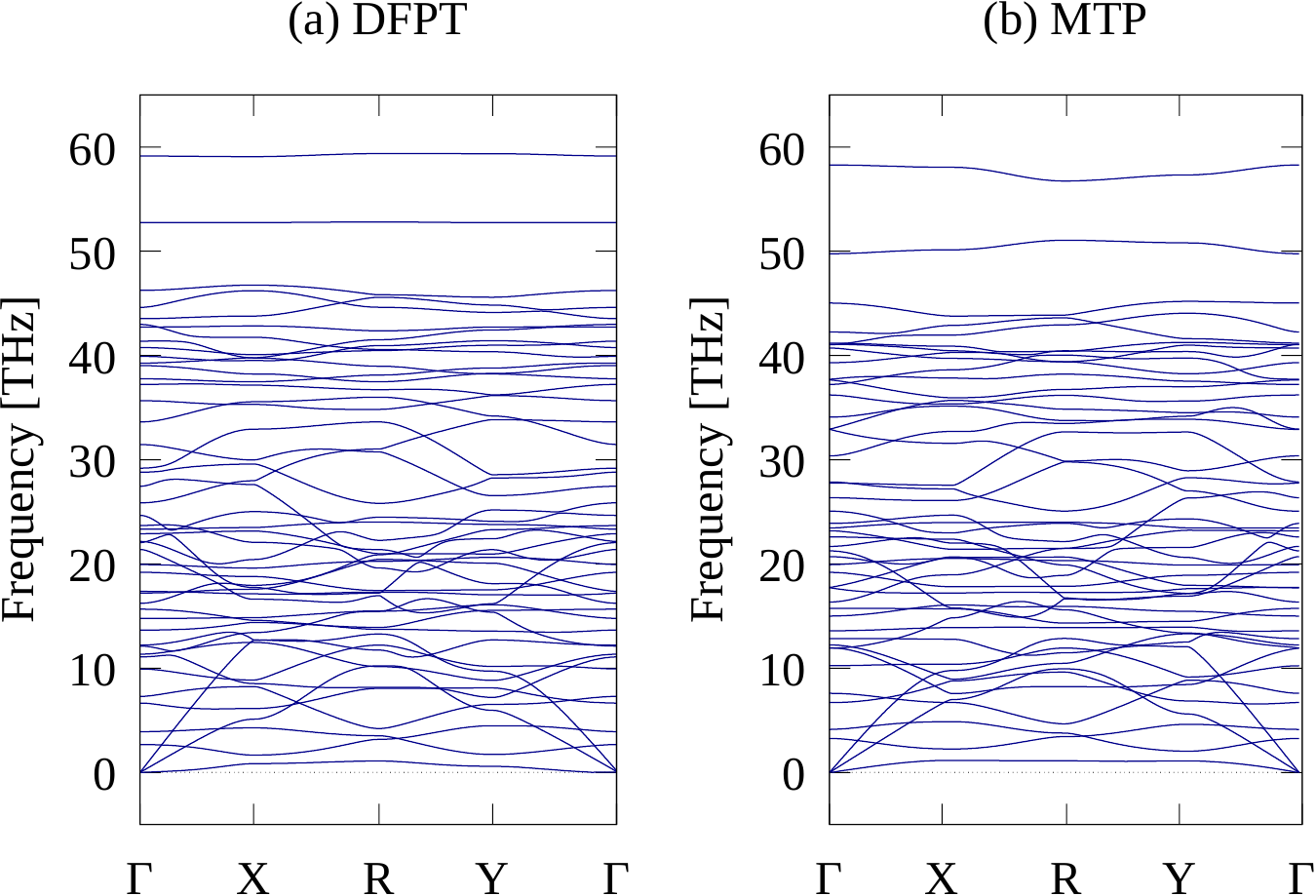}
	\caption{Phonon dispersion spectrum of the C$_{16}$ monolayer obtained using (a) DFTP and (b) MTP.}
	\label{fig:phonon}
\end{figure*}

\begin{figure}[!htb]
	\centering
	\includegraphics[width=1.0\linewidth]{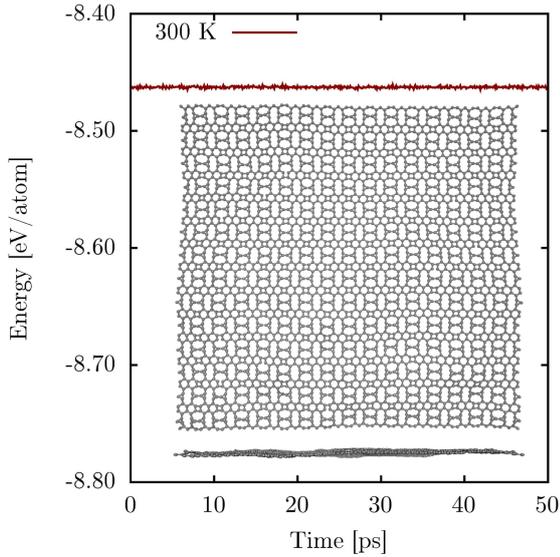}
	\caption{C$_{16}$ total energy values per atom as a function of the AIMD simulation. The insets show the top and lateral views of the final snapshot.}
	\label{fig:aimd}
\end{figure}

To determine the C$_{16}$ stability, we have calculated the cohesive energy value, $E_\text{coh/atom} = (E_\text{tot} - 16\cdot E_\text{C})/16 = $ \SI{-7.1}{eV/atom}. The dynamical structural stability was verified by calculating the phonon dispersion band structure using density functional perturbation theory (DFPT) and MTP. The results are consistent and presented in Fig.~\ref{fig:phonon}. There is a close similarity in the acoustic modes. The differences in the optical modes are more pronounced. These results validate the trained MLIP. Additionally, the absence of imaginary frequencies suggests that C$_{16}$ is structurally stable. There are two flat modes, around \SI{54}{THz} and \SI{59}{THz}. The lack of vibrational modes with frequencies exceeding \SI{60}{THz} are consistent with the sp$^2$ hybridization since very high vibrations are typically associated with an sp hybridization.

To further verify the system's stability, Fig.~\ref{fig:aimd} illustrates the total energy values temporal evolution of C$_{16}$ from an AIMD simulation carried out at a thermal bath temperature of \SI{300}{\kelvin} for over a \SI{5}{ps} simulation time. The results show that the system's total energy is conserved, as expected. The figure's inset presents C$_{16}$ snapshots from front and side views, which display no signs of atomic reconfiguration, bond rearrangement, or bond breaking. Furthermore, the side view reveals no significant fluctuations other than those expected from temperature variations.

\subsection{Electronic, Excitonic, and Optical Properties}

\begin{figure*}[!htb]
	\centering
	\includegraphics[width=0.7\linewidth]{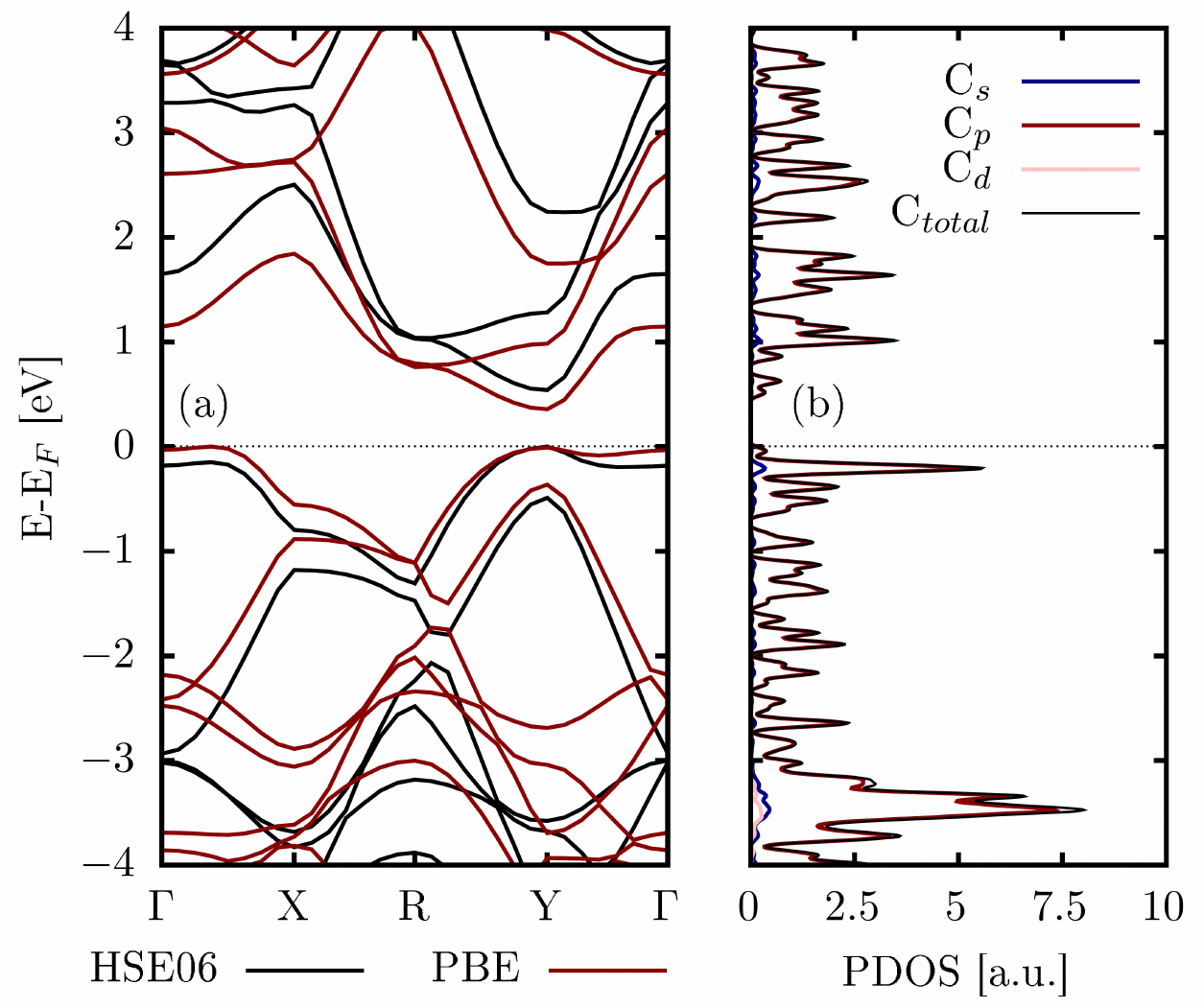}
	\caption{(a) C$_{16}$ electronic band Structure at the PBE (red curves) and HSE06 (black curves) levels and (b) the HSE06 Projected Density of States on the $s$-, $p$-, and $d$-orbitals. The Fermi level is set at \SI{0}{\electronvolt}.}
	\label{fig:bands_dos}
\end{figure*}

The electronic band structure and the corresponding projected density of states (PDOS) are presented in Fig.~\ref{fig:bands_dos} (a)-(b), respectively. Our system has a direct HSE06 (PBE) band gap of \SI{0.59}{\electronvolt} (\SI{0.36}{\electronvolt}) localized close to the Y high-symmetry point within the Y-$\Gamma$ path. From the PDOS, we can observe that the main orbital contributions around the Fermi level come from the \ce{C} $p$-orbitals, complemented by a minimal contribution of the \ce{C} $s$-orbitals within the region close to the valence band maximum (VBM).

\begin{figure*}[!htb]
    \centering
    \includegraphics[width=0.9\linewidth]{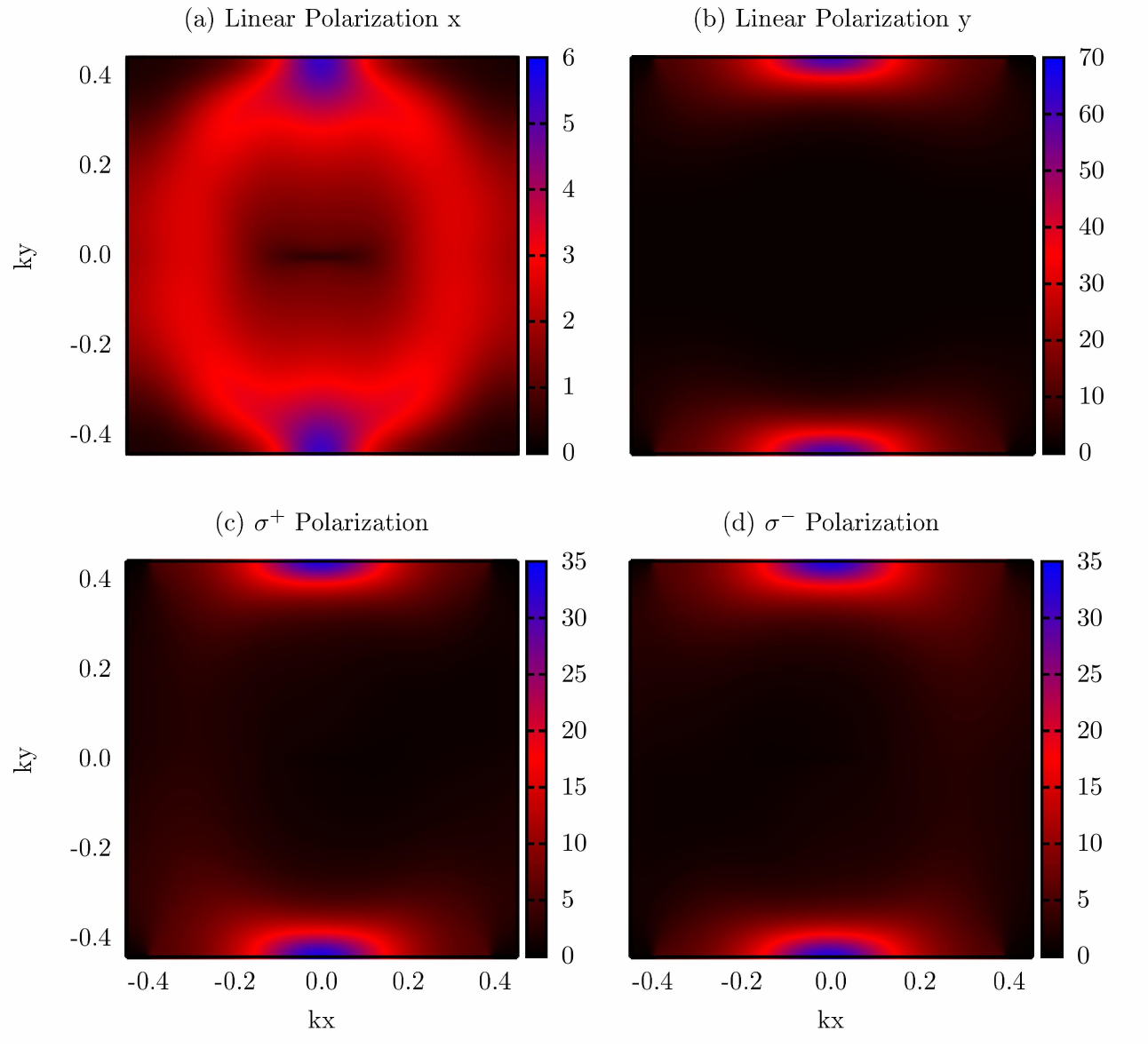}
    \caption{The optical activity of \ce{C16} in the first Brillouin zone considering linear (a,b) and circular polarized light (c,d).}
    \label{fig:optical_activity}
\end{figure*}

The optical activity, derived from the sum of the overall optical transitions by considering the two highest valence bands and the three lowest conduction bands, was analyzed for linear and circularly polarized light at the IPA level. This information is illustrated in Fig.~\ref{fig:optical_activity} (a)-(d). For linear polarized light along the $\hat{x}$ direction (Fig.~\ref{fig:optical_activity} (a)) we have some optical activity in the entire Brillouin zone (BZ), with higher values at the BZ edges at $k_x$=\SI{0}{\per\angstrom} and $k_y$= $\pm$ \SI{0.4}{\per\angstrom}, respectively, with a maximum oscillator force of \SI{6}{\square\angstrom}. When comparing the optical activity for the $\hat{y}$ polarization, shown in Fig.~\ref{fig:optical_activity} (b), the maximum values are in the same region of the previous panel; however, the oscillator force values are 10 times higher ($\sim$ \SI{70}{\square\angstrom}), with the optical activity concentrated at the BZ edges. These results show a large optical anisotropy, with the optical response much more intense along the  $\hat{y}$ direction. 

Fig.~\ref{fig:optical_activity} (c)-(d) illustrate the optical activity for circular light polarization, obtained from a linear combination of $\hat{x}$ and $\hat{y}$ linear polarizations. The isotropic behavior shown in this case results from the dominant optical activity for the $\hat{y}$ polarization, as the $\hat{x}$ contribution is much smaller, resulting in practically the same linear optical response under optical helicity.

\begin{figure}[!h]
	\centering
	\includegraphics[width=0.9\linewidth]{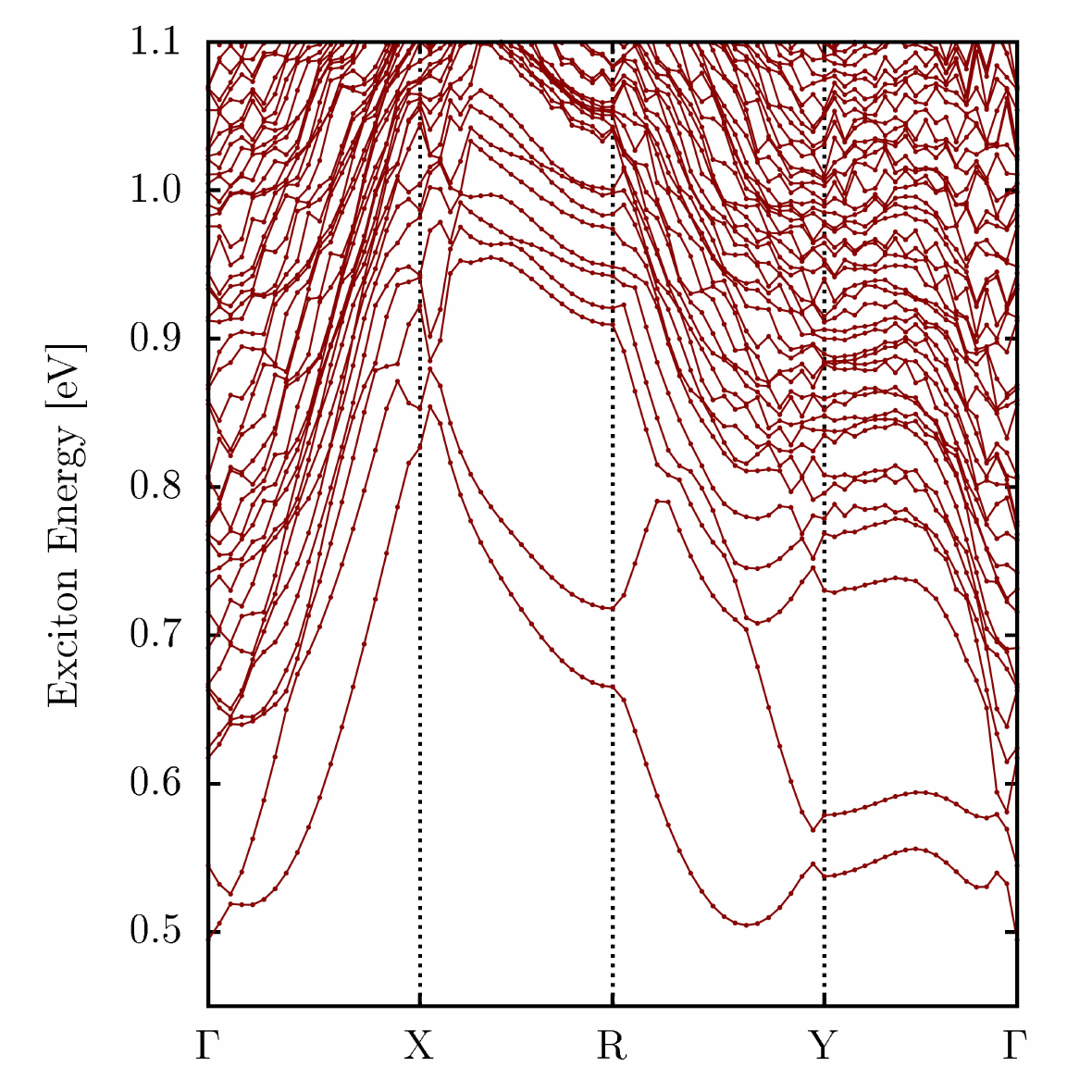}
	\caption{C$_{16}$ excitonic band structure obtained using an HSE06 parametrization in a MLWF-TB model.}
	\label{fig:exc_bands}
\end{figure}

The lowest energy excitonic levels, for direct (at $\Gamma$) and indirect (any other \textbf{k}-points) excitons, are shown in Fig.~\ref{fig:exc_bands}, the $x$-axis representing the excitonic momentum along the \textbf{k}-path, passing through the high symmetry points in the BZ. The excitonic ground state is direct, evidenced by the direct band gap observed in the electronic band structure. When compared with the fundamental electronic band gap, it results in an exciton binding energy of \SI{96}{\milli\electronvolt}, which is higher if compared to bulk values but lies within the lower limit for 2D carbon allotropes, where the expected exciton binding energy resides in the range [\SIrange{80}{500}{\milli\electronvolt}].\cite{Dias_2024}

\begin{figure*}[!htb]
	\centering
	\includegraphics[width=0.9\linewidth]{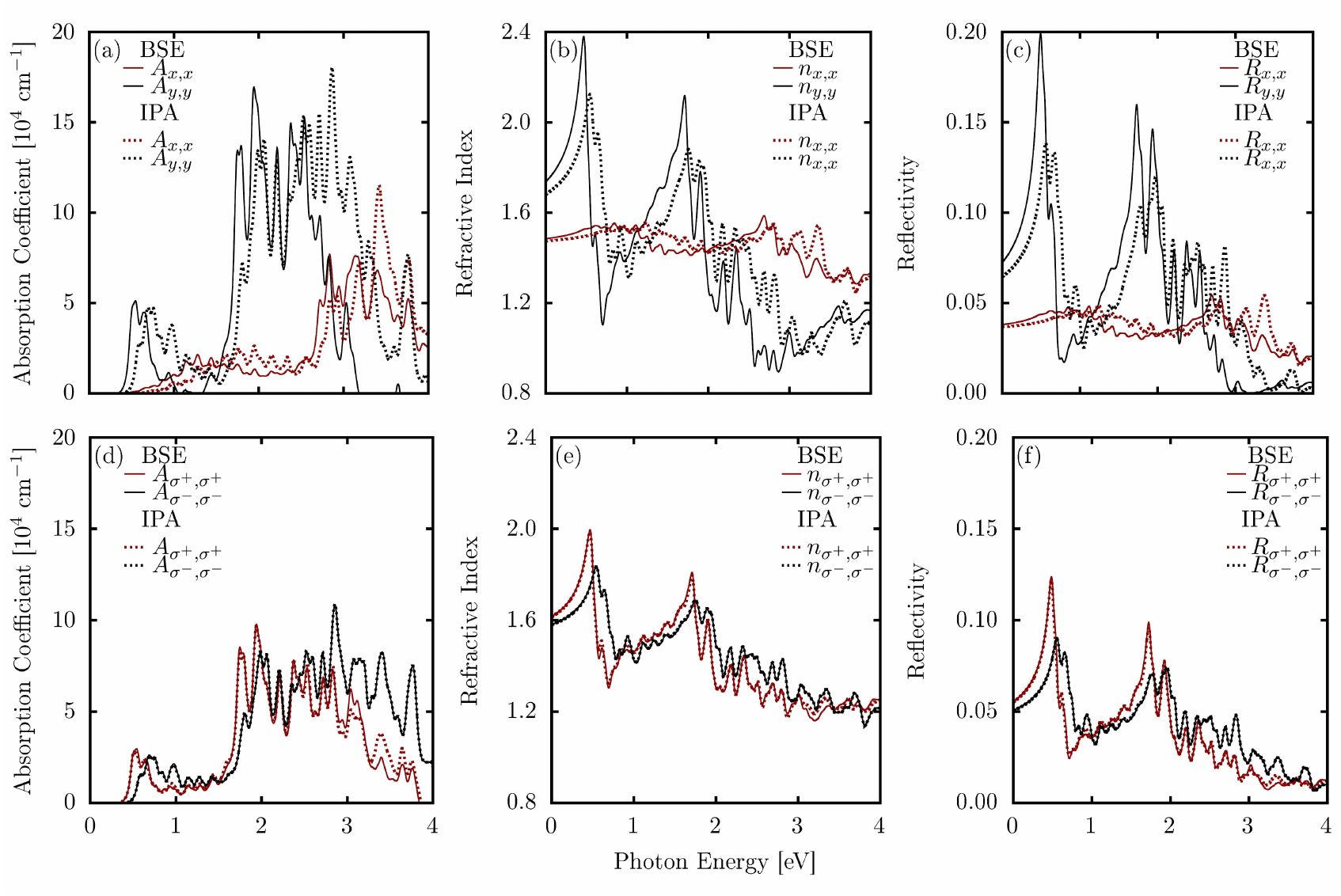}
	\caption{(a,d) Absorption Coefficient, (b,e) Refractive Index, and (c,f) Reflectivity, calculated within the BSE (solid curves) and IPA (dashed curves) approximations for linear (a-c) and circular light polarization (d-f). Red (black) curves correspond to $\hat{x}$ ($\sigma^{+}$) and $\hat{y}$ ($\sigma^{-}$) linearly (circularly) polarized light cases.}
	\label{fig:optical_properties}
\end{figure*}

The linear optical response, at the IPA and BSE levels considering linear and circular light polarizations, through the evaluation of the absorption coefficient, refractive index, and reflectivity is shown in Fig.~\ref{fig:optical_properties}(a)-(f). Fig.~\ref{fig:optical_properties} (a) and (d)) demonstrate that C$_{16}$ excitonic band structure obtained using an HSE06 parametrization in an MLWF-TB model absorbs in the infrared, visible, and ultraviolet regions. The excitonic effects result in a slight red shift in the absorption's peaks. We also observe a substantial change in the linear optical response, giving rise to an important optical anisotropy for linear light polarization because the absorption coefficient is much more intense for light polarization along the $\hat{y}$ direction. This result is consistent with the optical activity shown in Fig.~\ref{fig:optical_activity}. In contrast, for light excitations above \SI{3.0}{\electronvolt}, the absorption coefficient is much more intense for $\hat{x}$ light polarization. The optical response is isotropic for circular light polarization for the above-mentioned reasons.

The optical anisotropy, under linear light polarization and the isotropic behavior for the circular one, is also seen in the refractive index (Fig.~\ref{fig:optical_properties} (b) and (e)) and the reflectivity (Fig.~\ref{fig:optical_properties} (c) and (f)). Under linear light polarization, the refractive index is higher for the $\hat{y}$ polarization in the infrared and visible spectrum regions. In the ultraviolet region, it is more pronounced for the $\hat{x}$ polarization. This index varies in the range of \SIrange{0.9}{2.4}{} and \SIrange{1.28}{2.4}{} for the $\hat{y}$ and  $\hat{x}$ polarizations, respectively. This range is slightly narrowed when excitonic effects are considered (BSE level). 

This behavior is similar to reflectivity that lies in the range of \SIrange{0}{20}{\percent} for the $\hat{y}$ polarization case and \SIrange{3}{5}{\percent} for the  $\hat{x}$ polarization case. It is also interesting to observe that none of the incident photons are reflected for $\hat{y}$ polarization in the ultraviolet region. When circular light polarizations are considered, both refractive index and reflectivity are the same, independent of clockwise ($\sigma_{-}$) or counterclockwise ($\sigma_{+}$) light polarization.

We can estimate the C$_{16}$  solar harvesting efficiency through the electronic and optical properties data, using the power conversion efficiency (PCE) as a descriptor. The PCE was obtained by two methods: the Shockley--Queisser limit (SQ) \cite{Shockley_510_1961} and the spectroscopy limited maximum efficiency (SLME).\cite{Yu_068701_2012} The former is directly obtained by the fundamental electronic band gap $E_{g}$ when excitonic effects are not considered or through the exciton ground state energy $Ex_{gs}$; this kind of approach considers that all photons with energies equal or higher than $E_{g}$ or $Ex_{gs}$ are absorbed. The latter approach, which goes beyond $E_{g}$ and $Ex_{gs}$, also considers the optical band gap or the exciton bright ground state energy when the excitonic effects are taken into account. Here, the absorption spectrum and the material thickness are the two fundamental parameters needed to estimate the optical absorbance that provides the incident photon absorption rate. For 2D materials, the value of van der Waals length (\SI{3.21}{\angstrom}) is added. This procedure is based on the work of Bernardi \textit{et al.},\cite{Bernardi_3664_2013} where it was used to estimate graphene's absorbance due to its atomic thickness, providing results consistent with experimental data. For both approaches, the PCE was also estimated considering the solar cell at \SI{300}{\kelvin} and considering the AM1.5G model for the solar emission spectrum.\cite{Astm_1_2012} It is equally important to understand that both methods provide an upper limit of the solar harvesting efficiency and that experimentally reaching these values can demand years of research. Details of the theoretical formalism behind both methods can be found in previous works.\cite{Moujaes_111573_2023,Dias_108636_2023}

The SQ-limit estimates a PCE$^{SQ}$ of \SI{17.06}{\percent} when excitonic effects are not considered; yet, when quasiparticle effects are taken into account, the value is reduced to \SI{12.96}{\percent},  due to the exciton binding energy, which results in $Ex_{gs}$, being smaller than $E_{g}$. The SQ-Limit curve has a maximum PCE$^{SQ}$ when either $E_{g}$ or $Ex_{gs}$ comes closer to \SI{1.31}{\electronvolt}; the further $E_{g}$ or $Ex_{gs}$ are from this ideal value, the lower is PCE$^{SQ}$. The SLME values (PCE$^{SLME}$) are significantly lower than those obtained through the previous approximation, resulting in a solar harvesting efficiency of less than \SI{0.08}{\percent}, regardless of whether quasi-particle effects are present. This low efficiency is primarily due to the material's monolayer thickness, which leads to a poor photon absorbance rate, as most incident photons pass through the material without being absorbed. Consequently, the material is almost transparent, presenting a considerable challenge for the use of 2D monolayers in photovoltaic devices.

Based on the investigation of Jariwala and co-workers,\cite{Jariwala_2962_2017} which proposes a solution for this problem using light trapping techniques to enhance the absorbance rate closer to \SI{100}{\percent} in 2D materials, we made another estimation of PCE (PCE$^{SLME}_{max}$), adapting the SLME method, and considering the absorbance rate as a Heaviside function, where we adopt the value of $1$ for incident photons with energy equals or higher than the optical band gap or exciton bright ground state and $0$ otherwise. PCE$^{SLME}_{max}$ evaluates to an efficiency of \SI{12.96}{\percent} with excitonic effects considered and \SI{17.05}{\percent} when they are absent, suggesting a potential application of this monolayer as a solar harvesting absolver if light trapping schemes are applied. 

The above results are consistent with previous investigations of 2D carbon allotropes for photovoltaic applications.\cite{Dias_2024} It is crucial to report the performance characteristics (PCE) both with and without excitonic effects, as the exciton binding energy is highly sensitive to the surrounding dielectric environment. Excitonic effects tend to be reduced by proximity interactions when a monolayer is in contact with other materials. This scenario is particularly relevant in solar cells, where the absorber layer is sandwiched between the electron and hole transport layers, leading to experimentally lower excitonic effects. In this work, we have explored both extremes: the maximum excitonic effects, which occur when the monolayer is in a vacuum, and the complete absence of these quasi-particle effects, which provide a more realistic value for the PCE. The actual performance is expected to fall somewhere between these two extremes.

\subsection{Thermoelectric Properties} 

The effectiveness of a material in thermoelectric devices is measured by probing its figure of merit (zT) \cite{goldsmid2010introduction,snyder2008complex} defined as: 
\begin{equation*}
zT=\frac{S^2\sigma}{\kappa_e+\kappa_L}~,
\end{equation*}
where $S$ is the Seebeck coefficient and $\sigma$ the electronic conductivity; $\kappa_e$ and $\kappa_L$ are the thermal electronic and thermal lattice conductivities, respectively. The quantity $S^2\sigma$ is the so-called power factor (PF); it represents the ability to produce electrical power from temperature gradients. Within the Boltzmann transport theory,\cite{harris2004introduction,jones1985theoretical} the tensor components of these physical quantities, except $\kappa_L$, can be extracted by evaluating:

\begin{equation*}
\kappa_e^{\alpha\beta}(E)=\sum_{n,k}\tau_{n,k} v_g^{\alpha}(n,k)v_g^{\beta}(n,k)\delta(E-E_{n,k})
\end{equation*}
\begin{eqnarray*}
\sigma^{\alpha\beta}(T,\eta)&=&\frac{e^2}{NV}\int \sum_{n,k}\tau_{n,k} v_g^{\alpha}(n,k)v_g^{\beta}(n,k)\\
                    & &\times~\delta(E-E_{n,k})[-\frac{\partial f(T,E,\eta)}{\partial E}] dE
\end{eqnarray*}
\begin{eqnarray*}
S^{\alpha\beta}(T,\eta)&=&\frac{1}{eVT\sigma^{\alpha\beta}(T,\eta)}\int \kappa_{e}^{\alpha\beta}(E-\eta)\\
               & &\times~[-\frac{\partial f(T,E,\eta)}{\partial E}] dE~.
\end{eqnarray*}
$e$ is the electron's charge, $T$ is temperature, $E$ represents energy, $N$ is the number of electronic (\textbf{k}-points) sampled in the Brillouin zone, $V$ is the volume of the unit cell, and $\eta$ is the chemical potential. Moreover, $f$ is the Fermi-Dirac distribution function, $\delta$ is the Dirac delta function, E$_{n,k}$ is the energy of the $n^{\rm{th}}$ band at the $k^{\rm{th}}$ wave vector, and $\tau_{n,k}$ is the electron/hole lifetime at point ($n$,$k$) in the $E$-$k$ plane. Likewise, $v_g^{\alpha}$ is the $\alpha^{\rm{th}}$ component of the group velocity $v_g$. It should be noted that $S$ is independent of the electronic lifetime, whereas $\sigma$ and $\kappa_e$ are not.

\begin{figure*}[p]
	\centering
	\includegraphics[width=0.8\linewidth]{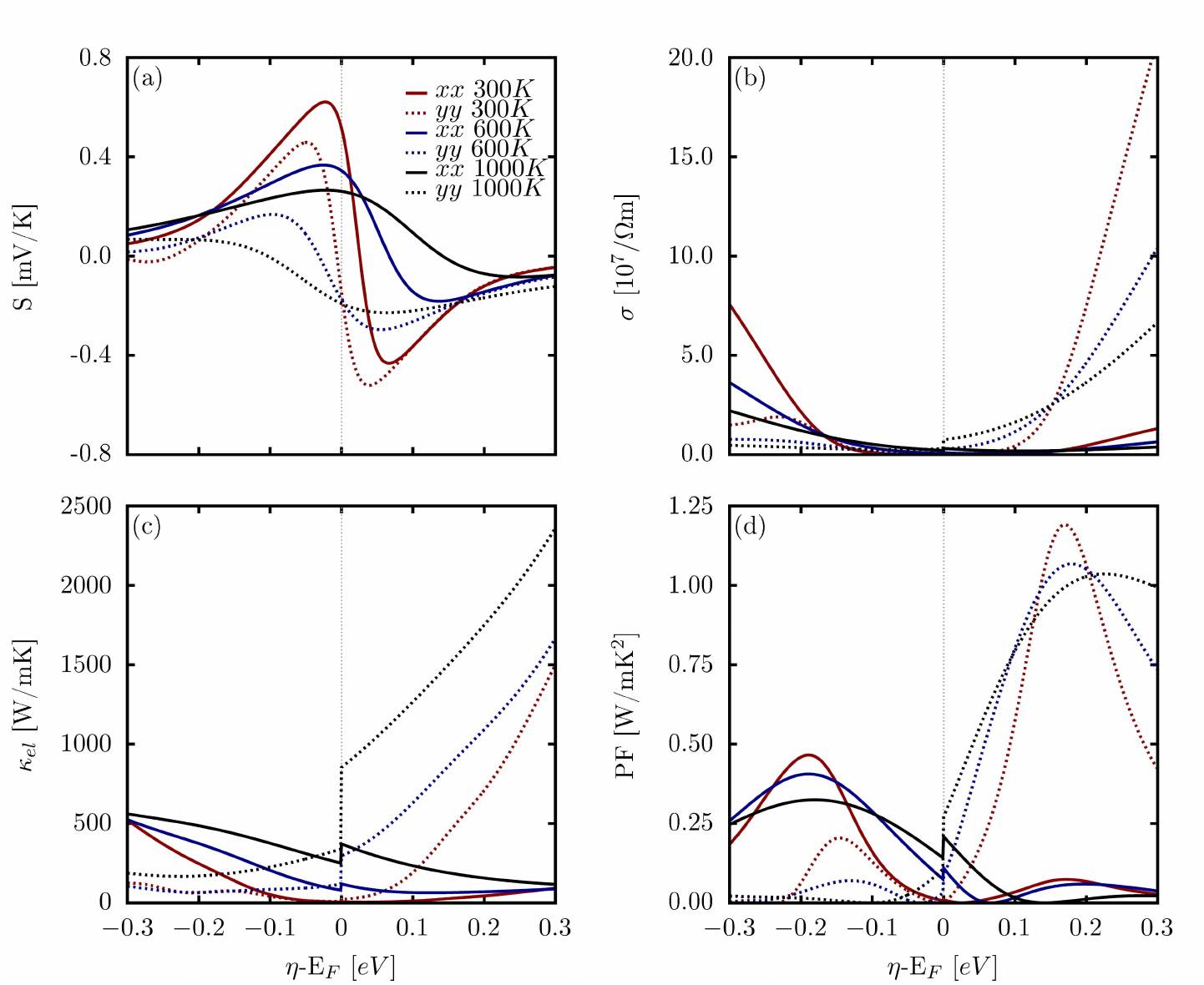}
	\caption{(a) The change of the Seebeck coefficient, (b) the electronic conductivity, (c) the thermal electronic conductivity, and (d) the power factor as a function of electron and hole doping along the $x$ and $y$ directions at \SI{300}{\kelvin} (red), \SI{600}{\kelvin} (blue), and \SI{1000}{\kelvin} (black).}
	\label{thermoelectric}
\end{figure*}

\begin{figure*}[!p]
	\centering
	\includegraphics[width=0.8\linewidth]{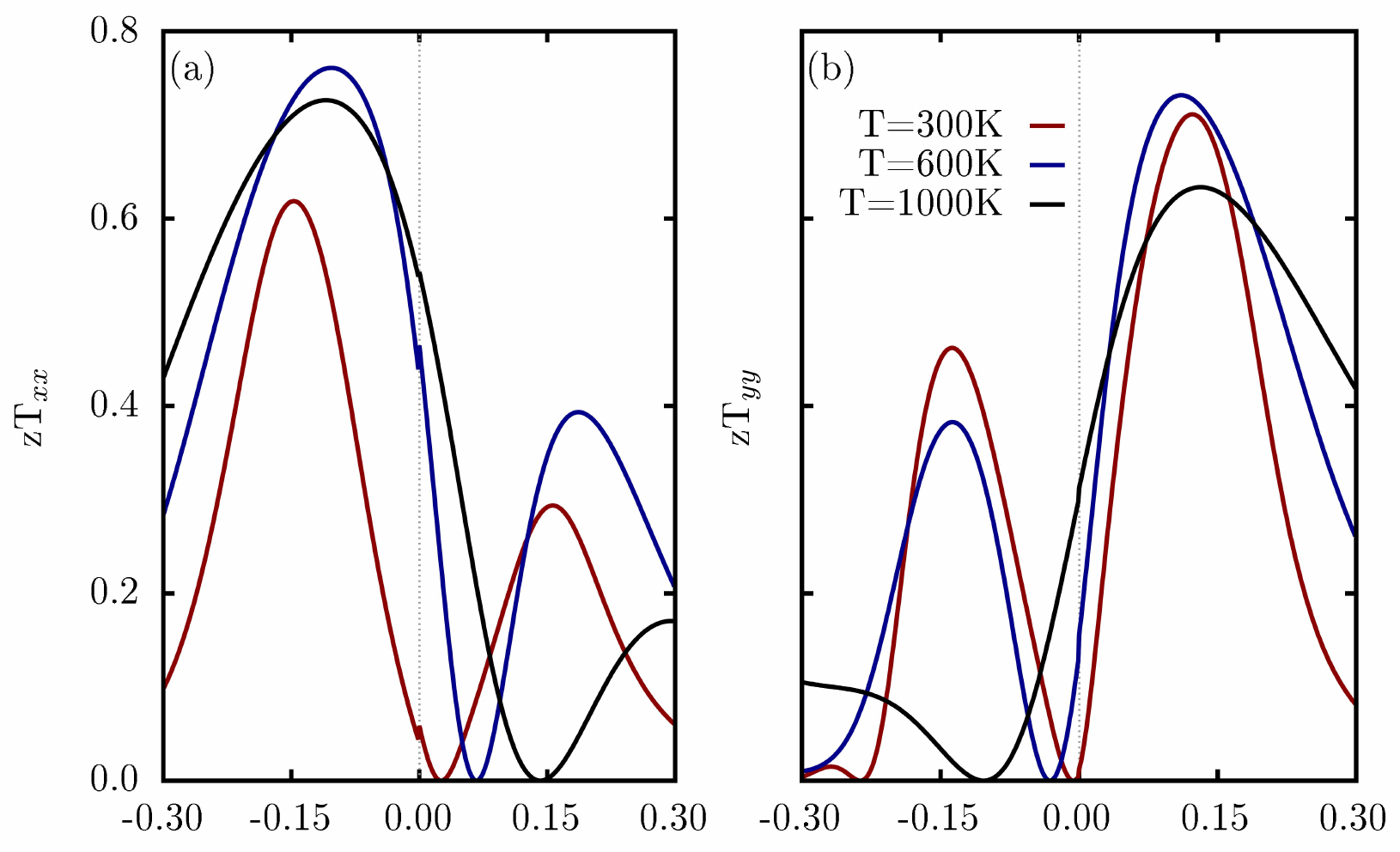}
	\caption{Same as in Fig. \ref{thermoelectric} but for the figure of merit (zT) at \SI{300}{\kelvin}, \SI{600}{\kelvin}, and \SI{1000}{\kelvin} along the (a) $x$ and (b) $y$ directions.}
	\label{fig:zT}
\end{figure*}

The computation of $\tau_{n,k}$ requires the knowledge of the imaginary part of the electron's self-energy, deduced by evaluating electron-phonon matrix elements on a highly dense electronic grid. Since this is a burdensome job and since one single value of $\tau$ per carrier charge is requisite for the BoltzWann code, the electronic lifetime will be directly obtained from the mobility values of the 2D C$_{16}$ structure ($\mu_{2D,i}$) via the expression:
\begin{equation*}
\tau_i=\frac{\mu_{2D,i}m_{i}^{*}}{e},
\end{equation*}
with $i$ referring to electrons ($e$) or holes ($h$). Along the $x$ and $y$ directions, $\mu_e >$ $\mu_h$ and $\tau_e > \tau_h$ at all temperatures.

\begin{table*}[]
\begin{tabular}{|cccccccc|}
\hline
\multicolumn{8}{|c|}{$T = 300$ K}                                                                                                                                                                                                                                                                                                  \\ \hline
\multicolumn{1}{|c|}{Direction}            & \multicolumn{1}{c|}{Carrier}      & \multicolumn{1}{c|}{$m_i^*$ ($m_0$)} & \multicolumn{1}{c|}{$m_d$ ($m_0$)} & \multicolumn{1}{c|}{$C_{2D}$ (eV/\r{A})} & \multicolumn{1}{c|}{$E_1$ (eV)} & \multicolumn{1}{c|}{$\mu$ (10$^3$cm$^2$V$^{-1}$s$^{-1}$)} & $\tau$ (ps) \\ \hline
\multicolumn{1}{|c|}{\multirow{2}{*}{$x$}} & \multicolumn{1}{c|}{\textit{e}}   & \multicolumn{1}{c|}{0.15}            & \multicolumn{1}{c|}{0.021}         & \multicolumn{1}{c|}{6.19}                                 & \multicolumn{1}{c|}{-9.95}      & \multicolumn{1}{c|}{61.2}                                 & 5.227       \\ \cline{2-8} 
\multicolumn{1}{|c|}{}                     & \multicolumn{1}{c|}{\textit{h}}   & \multicolumn{1}{c|}{0.29}            & \multicolumn{1}{c|}{0.081}         & \multicolumn{1}{c|}{6.19}                                 & \multicolumn{1}{c|}{-8.68}      & \multicolumn{1}{c|}{21.2}                                 & 3.500       \\ \hline
\multicolumn{1}{|c|}{\multirow{2}{*}{$y$}} & \multicolumn{1}{c|}{\textit{e}}   & \multicolumn{1}{c|}{0.14}            & \multicolumn{1}{c|}{0.021}         & \multicolumn{1}{c|}{7.95}                                 & \multicolumn{1}{c|}{6.80}       & \multicolumn{1}{c|}{180.4}                                & 14.380      \\ \cline{2-8} 
\multicolumn{1}{|c|}{}                     & \multicolumn{1}{c|}{\textit{h}}   & \multicolumn{1}{c|}{0.28}            & \multicolumn{1}{c|}{0.081}         & \multicolumn{1}{c|}{7.95}                                 & \multicolumn{1}{c|}{7.68}       & \multicolumn{1}{c|}{36.01}                                & 5.739       \\ \hline
\multicolumn{8}{|c|}{$T = 600$ K}                                                                                                                                                                                                                                                                                                  \\ \hline
\multicolumn{1}{|c|}{Direction}            & \multicolumn{1}{c|}{Carrier}      & \multicolumn{1}{c|}{$m_i^*$ ($m_0$)} & \multicolumn{1}{c|}{$m_d$ ($m_0$)} & \multicolumn{1}{c|}{$C_{2D}$ (eV/\r{A})} & \multicolumn{1}{c|}{$E_1$ (eV)} & \multicolumn{1}{c|}{$\mu$ (10$^3$cm$^2$V$^{-1}$s$^{-1}$)} & $\tau$ (ps) \\ \hline
\multicolumn{1}{|c|}{\multirow{2}{*}{$x$}} & \multicolumn{1}{c|}{\textit{e}}   & \multicolumn{1}{c|}{0.15}            & \multicolumn{1}{c|}{0.021}         & \multicolumn{1}{c|}{6.19}                                 & \multicolumn{1}{c|}{-9.95}      & \multicolumn{1}{c|}{30.6}                                 & 2.613       \\ \cline{2-8} 
\multicolumn{1}{|c|}{}                     & \multicolumn{1}{c|}{\textit{$y$}} & \multicolumn{1}{c|}{0.29}            & \multicolumn{1}{c|}{0.081}         & \multicolumn{1}{c|}{6.19}                                 & \multicolumn{1}{c|}{-8.68}      & \multicolumn{1}{c|}{10.5}                                 & 1.734       \\ \hline
\multicolumn{1}{|c|}{\multirow{2}{*}{$y$}} & \multicolumn{1}{c|}{\textit{e}}   & \multicolumn{1}{c|}{0.14}            & \multicolumn{1}{c|}{0.021}         & \multicolumn{1}{c|}{7.95}                                 & \multicolumn{1}{c|}{6.80}       & \multicolumn{1}{c|}{90.2}                                 & 7.190       \\ \cline{2-8} 
\multicolumn{1}{|c|}{}                     & \multicolumn{1}{c|}{\textit{h}}   & \multicolumn{1}{c|}{0.28}            & \multicolumn{1}{c|}{0.081}         & \multicolumn{1}{c|}{7.95}                                 & \multicolumn{1}{c|}{7.68}       & \multicolumn{1}{c|}{18.0}                                 & 2.869       \\ \hline
\multicolumn{8}{|c|}{$T = 1000$ K}                                                                                                                                                                                                                                                                                                 \\ \hline
\multicolumn{1}{|c|}{Direction}            & \multicolumn{1}{c|}{Carrier}      & \multicolumn{1}{c|}{$m_i^*$ ($m_0$)} & \multicolumn{1}{c|}{$m_d$ ($m_0$)} & \multicolumn{1}{c|}{$C_{2D}$ (eV/\r{A})} & \multicolumn{1}{c|}{$E_1$ (eV)} & \multicolumn{1}{c|}{$\mu$ (10$^3$cm$^2$V$^{-1}$s$^{-1}$)} & $\tau$ (ps) \\ \hline
\multicolumn{1}{|c|}{\multirow{2}{*}{$x$}} & \multicolumn{1}{c|}{\textit{e}}   & \multicolumn{1}{c|}{0.15}            & \multicolumn{1}{c|}{0.021}         & \multicolumn{1}{c|}{6.19}                                 & \multicolumn{1}{c|}{-9.95}      & \multicolumn{1}{c|}{18.4}                                 & 1.571       \\ \cline{2-8} 
\multicolumn{1}{|c|}{}                     & \multicolumn{1}{c|}{\textit{$y$}} & \multicolumn{1}{c|}{0.29}            & \multicolumn{1}{c|}{0.081}         & \multicolumn{1}{c|}{6.19}                                 & \multicolumn{1}{c|}{-8.68}      & \multicolumn{1}{c|}{6.4}                                  & 1.05        \\ \hline
\multicolumn{1}{|c|}{\multirow{2}{*}{$y$}} & \multicolumn{1}{c|}{\textit{e}}   & \multicolumn{1}{c|}{0.14}            & \multicolumn{1}{c|}{0.021}         & \multicolumn{1}{c|}{7.95}                                 & \multicolumn{1}{c|}{6.80}       & \multicolumn{1}{c|}{54.1}                                 & 4.312       \\ \cline{2-8} 
\multicolumn{1}{|c|}{}                     & \multicolumn{1}{c|}{\textit{h}}   & \multicolumn{1}{c|}{0.28}            & \multicolumn{1}{c|}{0.081}         & \multicolumn{1}{c|}{7.95}                                 & \multicolumn{1}{c|}{7.68}       & \multicolumn{1}{c|}{10.8}                                 & 1.722       \\ \hline
\end{tabular}
\caption{Calculated effective mass $m_i^*$, average effective mass $m_d$, in-plane stiffness $C_{2D}$, DP constant $E_1$, the charge carrier mobility $\mu$, and the electronic lifetime $\tau$ at $T=300$ K, 600 K, and 1000 K for the C$_{16}$. Here, $m_0$ is the  mass of a free electron.}
\label{mobility}
\end{table*}

\begin{figure}[t]
    \centering
    \includegraphics[width=8cm]{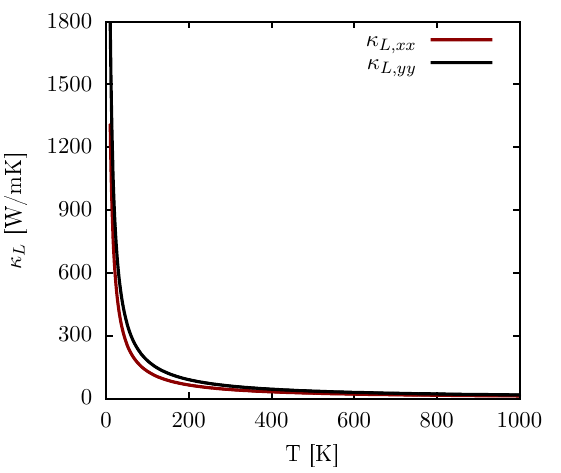}
    \caption{Lattice thermal conductivity along the $x$- (red) and $y$-directions (black).}
    \label{fig:kappaL}
\end{figure}

On the other hand, $\kappa_L$ is determined via:
\begin{equation*}
\kappa_L^{\alpha \beta}=\frac{1}{NV}\sum_{\textbf{q},s} c_v(\textbf{q},s)
 v_g^{\alpha}(\textbf{q},s) v_g^{\beta}(\textbf{q},s)\tau(\textbf{q},s)~,    
\end{equation*}
where $c_v$ is the phonon specific heat capacity and $\tau(\textbf{q},s)$ is the phonon lifetime of vibrational mode $s$ at the $\textbf{q}^{\rm{th}}$ (phonon) point. This calculation can be performed using the PHONO3PY code \cite{togo2015distributions,togo2023implementation}, which requires the tedious evaluation of third-order anharmonic constants. Such estimation is computationally very heavy, especially when the unit cell comprises many atoms. 

As an alternative, and assuming that the acoustic phonons only participate in the heat transport in the structure, an empirical expression of $\kappa_L$, derived from the Slack model, could be employed: \cite{morelli2006high, slack1965thermal, slack1973nonmetallic, raval2024novel}  
\begin{equation*}
\kappa_L=A\frac{\Bar{M}V^{1/3}\theta_{DA}^3}{\gamma^2 n^{2/3}T}~.   
\end{equation*}
Here, $\Bar{M}$ is the average atomic mass, $n$ is the number of atoms in the unit cell, $T$ is temperature, and $\theta_{DA}$ is the Debye temperature of the acoustic phonon modes.

$\gamma$ is the Gr\"{u}neisen parameter, which could be obtained from the PHONOPY code by running three phonon calculations, the first at the equilibrium volume and the others at slightly smaller and slightly larger volumes than the equilibrium one. Instead, it could be easily determined via an empirical expression that depends on the Poisson ratio ($\nu$) of the material: \cite{sanditov2009gruneisen, raval2024novel}
\begin{equation*}
\gamma=\frac{3}{2}\Big(\frac{1+\nu}{2-3\nu}\Big)~.
\end{equation*}
Since $\nu$ can be computed along any crystallographic direction, $\gamma$ becomes direction dependent too. On the other hand, $A$ is a coefficient inherent in the Slack expression and dependent on $\gamma$ such that: \cite{peng2016thermal, julian1965theory, raval2024novel} 
\begin{equation*}
A=\frac{2.4 \times 10^{-6}}{1-\frac{0.514}{\gamma}+\frac{0.228}{\gamma^2}}.    
\end{equation*}
It has been observed that such an expression overestimates the values of $\kappa_L$ in some materials; \cite{qin2022high} however, this is currently our best approach until an experimental value of $\kappa_L$ for the C$_{16}$ monolayer becomes available.

Fig.~\ref{thermoelectric} illustrates C$_{16}$ $S$, $\sigma$, $\kappa_e$, and PF along the $x$ and $y$ directions for $T=300$ \si{\kelvin}, \SI{600}{\kelvin}, and \SI{1000}{\kelvin}. These quantities are plotted for chemical potential values \SI{0.3}{\electronvolt} above and below the Fermi level to model electron and hole doping. For any temperature $T$, $S$ is larger along the $x$ direction. Except for $S$, the curves show some minor discontinuity when $\eta $ approaches $E_F$ from both sides due to the different electron and hole electronic lifetime values $\tau_e$ and $\tau_h$ (table \ref{mobility}) as $\eta-E_F\rightarrow 0^{+}$ and $\eta-E_F\rightarrow$0$^{-}$.

For the undoped C$_{16}$, that is for $\eta\sim E_F$ and $T=300$ K, $\kappa_e=6.05$ \si{W/(mK)} along the $x$ direction and $\kappa_e=22.79$ \si{W/(mK)} along the $y$-direction. These values increase as $T$ increases, with the electronic thermal conductivity along the $x$ direction always smaller than that along the $ y$ one. More specifically, $\kappa_{e, xx}=119.79$ \si{W/(mK)} and $\kappa_{e, yy}=296.87$ \si{W/(mK)} at $T=600$ \si{\kelvin}, and $\kappa_{e, xx}=371.14$ \si{W/(mK)} and $\kappa_{e, yy}=853.22$ \si{W/(mK)} at $T=1000$ \si{\kelvin}.

Regarding the PF, we also observed that the values along the $x$ direction are larger than those along the $y$ one, at least for $T=300$ and 600 \si{\kelvin}. At ambient temperatures, PF$_{xx}=9.70$ mW/(mK$^2$) while PF$_{yy}=3.36$ mW/(mK$^2$). At \SI{600}{\kelvin}, PF$_{xx}$ reaches a value of 0.11 W/(mK$^2$) and PF$_{yy}$ attains a value of 0.08 W/(mK$^2$).

As can be seen from Fig.~\ref{fig:kappaL}, the lattice thermal conductivity $\kappa_L$ shows smaller values along the $x$ direction. At $T=300$ \si{\kelvin} and for the undoped structure, $\kappa_{L,xx}=43.67$ W/(mK) and $\kappa_{L,yy}=60.67$ W/(mK). As $T$ is increased, $\kappa_L$ decreases such that $\kappa_{L,xx}=21.84$ W/(mK), and $\kappa_{L,yy}=30.34$ W/(mK) at \SI{600}{\kelvin}, and $\kappa_{L,xx}=13.10$ W/(mK), and $\kappa_{L,yy}=18.20$ W/(mK) at \SI{1000}{\kelvin}.

To achieve a significant value of zT, a maximum value of PF and a minimum value of the sum of the thermal conductivities are needed. As expected, the figure of merit along the $x$ direction is higher. For the undoped structure, zT registers values of 0.06, 0.46, and 0.54 at $T=300$ \si{\kelvin}, \SI{600}{\kelvin}, and \SI{1000}{\kelvin}, respectively. 

Since the structure is thermally stable at \SI{1000}{\kelvin}, a value of $\mathrm{zT}=0.54$ is relatively high; however, having zT values larger than or equal to 0.8 is desirable. Upon doping, zT$_{xx}$ (zT$_{yy}$) reaches a maximum value of 0.62 (0.71) for energies \SI{0.15}{\electronvolt} (\SI{0.12}{\electronvolt}) below (above) E$_F$ at ambient temperature. For $T=600$ \si{\kelvin}, the maxima obtained values of 0.76 (0.73) for zT$_{xx}$ (zT$_{yy}$) occurring at \SI{0.1}{\electronvolt} (\SI{0.11}{\electronvolt}) below (above) E$_F$. At a higher temperature of \SI{1000}{\kelvin}, zT$_{xx}$ has a maximum of 0.73 for an energy of \SI{0.11}{\electronvolt} below E$_F$, while zT$_{yy}$ depicts a maximum of 0.63 for an energy of \SI{0.13}{\electronvolt} above E$_F$. 

Knowing that energies below the Fermi level correspond to hole doping while those above to electron doping and that the farther we are from E$_F$, the heavier the doping, our best compromise is to induce a $p$-doping process capable of securing a zT$_{xx}$ of 0.76 at $T=600$ \si{\kelvin}, and by lowering E$_F$ by just \SI{0.1}{\electronvolt}. Considering that E$_F$ is proportional to the electron/hole density ($n$), then we can assume that $\Delta E_F/E_{F0}=\Delta n/n_0$, where $E_{F0}$ and $n_0$ are the Fermi energy and electron/hole concentration for the undoped system, respectively; $\Delta E_F=E_F$-$E_{F0}$ and $\Delta n=n$-$n_0$, $E_F$ being the Fermi energy and $n$ the carrier concentration of the doped system. We show that $n$/$n_0=1.042$, implying that $n=1.042n_0$, or $n \sim n_0$. Thus, by just lightly doping C$_{16}$ with holes corresponding to a \SI{0.1}{\electronvolt} shift in the Fermi energy, we can reach a zT as high as 0.76 along the $x$ direction.

\subsection{Mechanical Properties} 

\begin{figure}[b]
    \centering
    \includegraphics[width=0.9\linewidth]{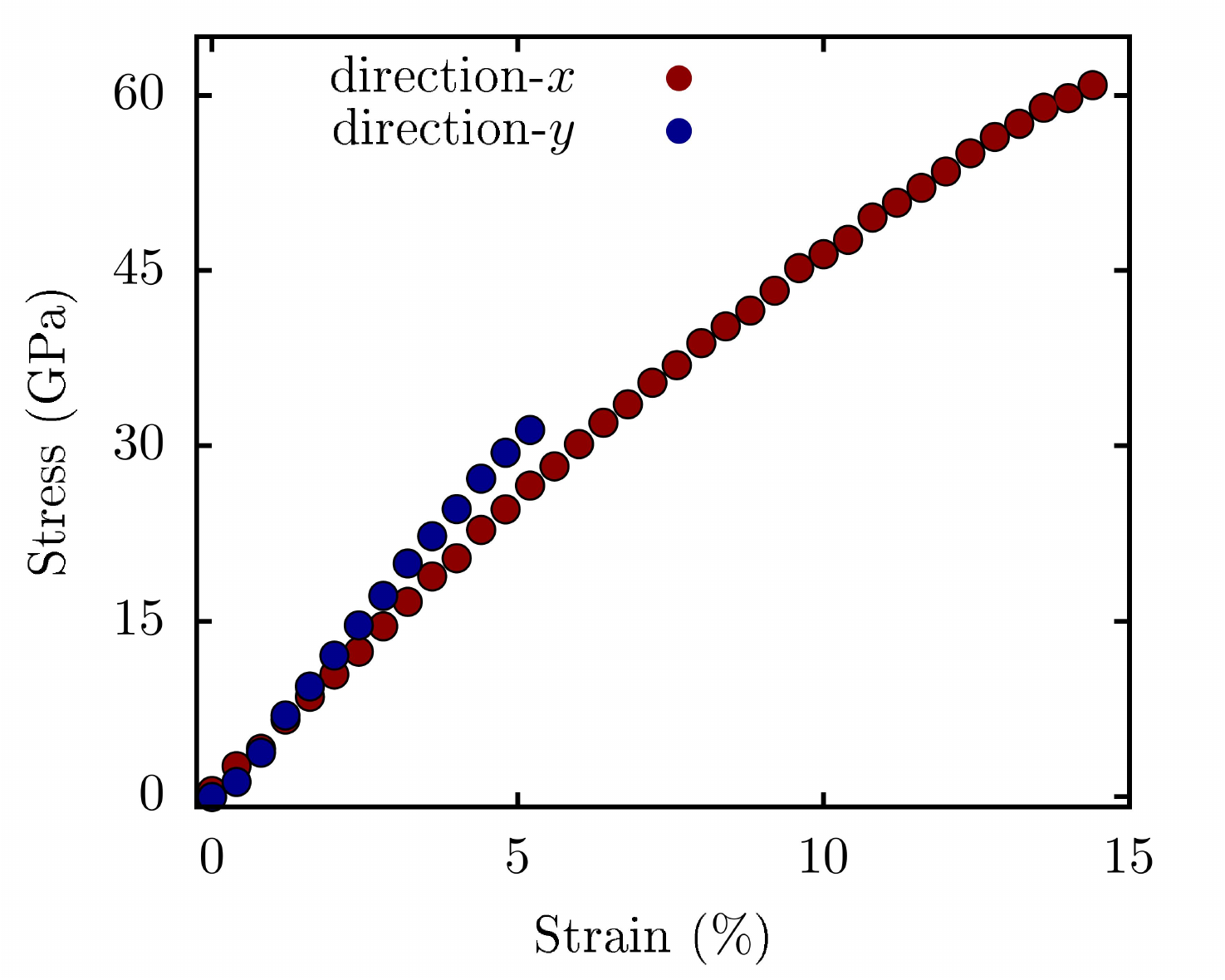}
    \caption{Stress-Strain curve, obtained using MLIP, for strain along the $x$- (red) and $y$-directions (blue).}
    \label{fig:stress-strain-Mlip}
\end{figure}

\begin{figure*}[!htb]
    \centering
    \includegraphics[width=0.9\linewidth]{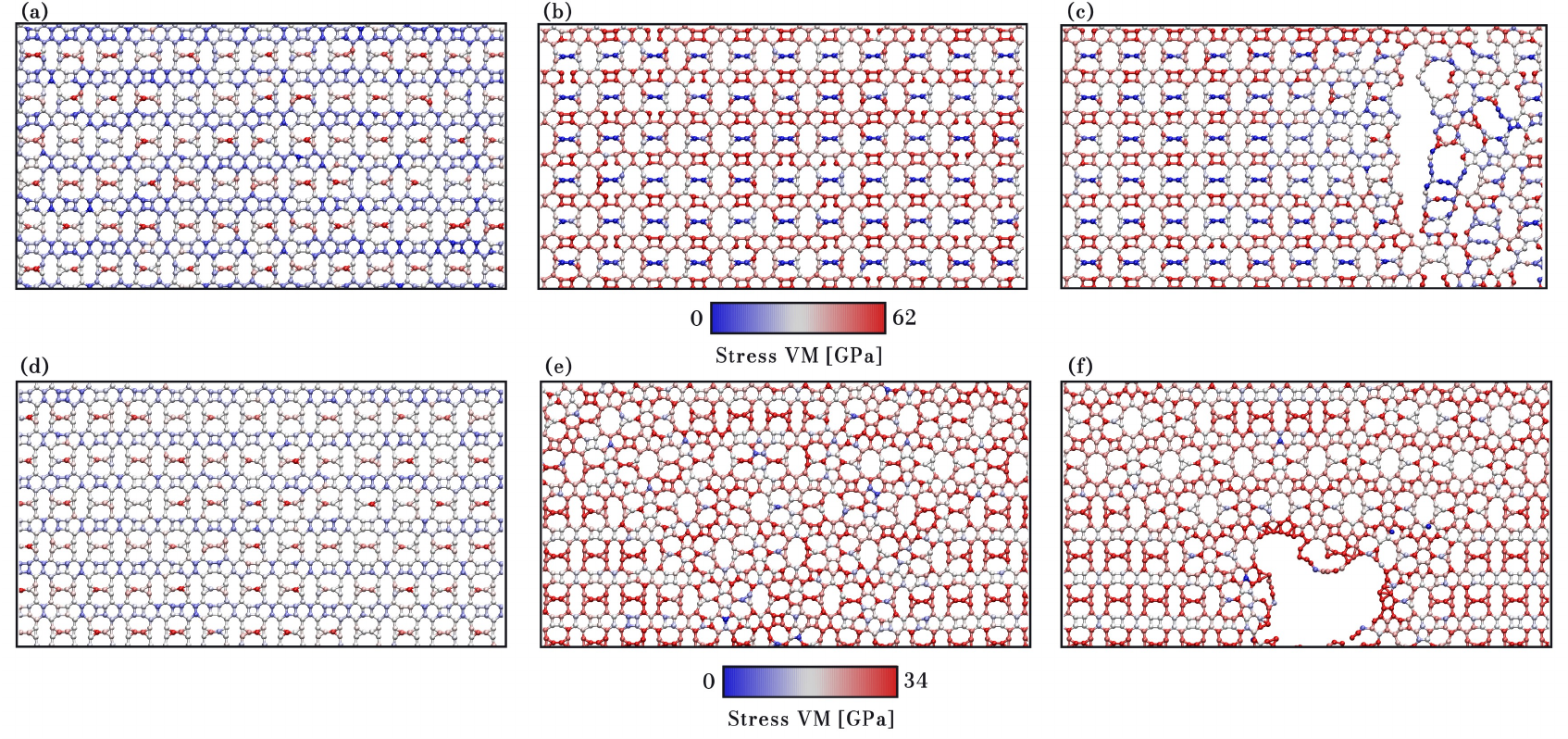}
    \caption{Representative snapshots of the C$_{16}$  monolayer at different stages of the deformation process along the $x$ (a-c) and $y$ (d-f) directions. The deformations shown are at (a) \SI{0}{\percent} , (b) \SI{14.4}{\percent}, and (c) \SI{14.7}{\percent} along the $x$ direction, and (d) \SI{0}{\percent}, (e) \SI{5.3}{\percent}, and (f) \SI{5.6}{\percent} along the $y$ one.}
    \label{fig:stress-strain-Painel}
\end{figure*}

To understand the C$_{16}$ mechanical response, we initially performed ab initio molecular dynamics (AIMD) simulations under various fixed strain values and with constant temperature variations. Based on these data, a machine learning interatomic potential (MLIP) was trained, which was then used to obtain the mechanical response of the system. A C$_{16}$ $16\times14\times1$ supercell containing 3584 atoms was subjected to deformations along the $x$ and $y$ directions up to a maximum of \SI{15}{\percent} of the respective initial length values. 

In Fig.~\ref{fig:stress-strain-Mlip}, we present the stress-strain curves obtained from these simulations. The red points represent the calculated stress for each deformation percentage along the $x$ direction. In contrast, the blue points show the corresponding results for uniaxial deformation applied along the $y$ direction. The system exhibits partially similar behaviors concerning hardness for these directions. Along the $x$ direction, Young’s modulus is approximately \SI{500}{\giga\pascal}, while along the $y$ one, this value reaches nearly \SI{630}{\giga\pascal}. Both cases display ductile characteristics; however, the stress accumulation along the $y$ direction is higher, leading the structure to fracture at a strain of only \SI{5.5}{\percent}. On the other hand, along the $x$ direction, the system endures nearly a three times greater deformation, with a critical strain at \SI{14.6}{\percent}. Given these aspects, the maximum stress obtained for deformation along the $x$ direction is approximately \SI{61.7}{\giga\pascal} at \SI{14.6}{\percent} strain and \SI{33.1}{\giga\pascal} along the $y$ one at \SI{5.5}{\percent} strain.

To gain deeper insights into the C$_{16}$ mechanical response, in Fig.~\ref{fig:stress-strain-Painel}, we present representative snapshots of C$_{16}$ under several strain values. Panels (a) and (d) depict the system's equilibrium state after residual stresses have been eliminated and the system thermalized at room temperature. The color gradient from blue to red represents the Von Mises (VM) stress \cite{mises_1913,pereira2020temperature,junior2020thermomechanical,pereira2022mechanical}, which provides a better understanding of where the fractures originate and propagate.

In panel (b), the system is shown at \SI{14.4}{\percent} strain on the limit of fracturing. The strain is concentrated in the region forming covalent chains of naphthalene, particularly in the rectangular rings, given their limited flexibility to release the strain. Finally, panel (c) shows the system at \SI{15}{\percent} strain, where the fracture occurs, and a crack propagates, with successive breakages of the hexagonal rings.

Along the $y$ direction, panel (e) shows the system under \SI{5}{\percent} strain. Several bonds have reconfigured, and it is possible to observe that stress is concentrated mainly in the regions derived from the bicyclopropylidene molecule. This region proved to be more fragile due to the topology of the involved rings, which explains the early system failure under strain along this direction. After the atomic reconfiguration upon fracture (panel (c) -- \SI{5.6}{\percent}), several bonds break, leading to the C$_{16}$ complete structural failure.

\section{Conclusion}

In summary, we propose a new two-dimensional carbon allotrope with a semiconductor behavior. We performed first-principles simulations based on the DFT formalism and atomistic simulations using a reactive force field trained through machine learning to characterize this novel nanomaterial. We conclude that the C$_{16}$ monolayer demonstrates promising physical and chemical properties, particularly in applications involving electronic devices and energy conversion. Electronically, 2D-C$_{16}$ displays a direct bandgap of \SI{0.59}{\electronvolt}, making it suitable for optoelectronic device applications. Its optical properties demonstrate anisotropic absorption along the $y$ direction, with the material possessing an exciton binding energy of \SI{96}{\milli\electronvolt}, confirming its potential in photovoltaic applications. The material holds a power conversion efficiency (PCE$^{SLME}_{max}$) of \SI{13}{\percent} when all photons are absorbed and excitonic effects are considered. The thermoelectric figure of merit ($zT$) can reach 0.8 at elevated temperatures, indicating that C$_{16}$ can be utilized in thermoelectric devices. Additionally, the anisotropic mechanical strength of C$_{16}$ is significant, with Young's moduli of \SI{500}{\giga\pascal} for the $x$ direction and \SI{630}{\giga\pascal} for the $y$ one, characterizing it as a ductile and robust material under mechanical stress.

\begin{acknowledgement}
The authors are thankful for financial support from the National Council for Scientific and Technological Development (CNPq, grant numbers $408144/2022-0$ and $305174/2023-1$), Federal District Research Support Foundation (FAPDF, grant numbers $00193-00001817/2023-43$ and $00193-00002073/2023-84$), the Coordination for Improvement of Higher Level Education (CAPES). In addition, the authors thank the ``Centro Nacional de Processamento de Alto Desempenho em S\~ao Paulo'' (CENAPAD-SP, UNICAMP/FINEP - MCTI project) for resources into the 897 and 570 projects, Lobo Carneiro HPC (NACAD) at the Federal University of Rio de Janeiro (UFRJ) for resources into 133 project. L.A.R.J acknowledges the financial support from FAPDF grant $0193.000942/2015$, CNPq grant $350176/2022-1$, and FAPDF-PRONEM grant $00193.00001247/2021-20$. A.C.D. and L.A.R.J also acknowledge PDPG-FAPDF-CAPES Centro-Oeste grant number $00193-00000867/2024-94$. M.L.P.J. acknowledges financial support from FAP-DF grant 00193-00001807/2023-16. Elie A. Moujaes would like to thank the financial support of the Brazilian National Council for Scientific and Technological Development CNPq, grant number $315324/2023-6$. K.A.L.L and D.S.G. acknowledge the Center for Computing in Engineering and Sciences at Unicamp for financial support through the FAPESP/CEPID Grant \#2013/08293-7.
\end{acknowledgement}

\bibliography{references.bib}

\end{document}